\documentclass{elsart}

\usepackage{epsfig,newlfont}

\DeclareMathAlphabet{\mathitbf}{T1}{cmr}{bx}{it}
\def\bbox#1{{ \mbox{\boldmath $#1$}}}
\def\bboxs#1{{ \mbox{\scriptsize\boldmath $#1$}}}

\usepackage{amssymb}
\usepackage{amsfonts,newlfont,epsfig}

\newcommand{\OO}[1]{\mathrm{O}(#1)}
\newcommand{\RP}[1]{\mathbf{R}\mathrm{P}^{#1}}

\newcommand{\s}{{\mathrm{s}}}
\newcommand{\f}{{\mathrm{f}}}
\newcommand{\V}{{\mathrm{V}}}
\newcommand{\ii}{{\mathrm{i}}}

\begin{document}
\begin{frontmatter}

\title{Numerical study of the enlarged O(5) symmetry of the 3-D
antiferromagnetic RP$^2$ spin model}

\author[ucm,bifi]{L.~A.~Fern\'andez,}
\author[ucm,bifi]{V.~Mart\'\i{}n-Mayor,} 
\author[unizar,bifi]{D.~Sciretti,}
\author[unizar,bifi]{A.~Taranc\'on,}
\author[unizar,bifi]{J.~L.~Velasco.}

\address[ucm]{Departamento de F\'{\i}sica Te\'orica, 
Facultad de Ciencias F\'{\i}sicas,\\
Universidad Complutense, 28040 Madrid, SPAIN}
\address[bifi]{Instituto de Biocomputaci\'on y  F\'{\i}sica de
Sistemas Complejos (BIFI)}
\address[unizar]{Departamento de F\'{\i}sica Te\'orica,
Facultad de Ciencias, \\
Universidad de Zaragoza, 50009 Zaragoza, SPAIN}

\date{\today}

\begin{abstract}
We investigate by means of Monte Carlo simulation and
Finite-Size Scaling analysis the critical properties of the three
dimensional $\OO5$ non linear $\sigma$ model and of the
antiferromagnetic $\RP2$ model, both of them regularized on a
lattice. High accuracy estimates are obtained for the critical
exponents, universal dimensionless quantities and critical
couplings. It is concluded that both models belong to the same
Universality Class, provide that rather non standard identifications
are made for the momentum-space propagator of the $\RP2$ model. We
have also investigated the phase diagram of the $\RP2$ model extended
by a second-neighbor interaction.  A rich phase diagram is found,
where most phase transitions are first order.
\end{abstract}

\begin{keyword}
Universality, Spin Models, Monte Carlo, Finite-Size Scaling.

\noindent{Pacs: 
64.60.Fr, 
05.10.Ln. 
}
\end{keyword}
\end{frontmatter}


\section{Introduction}

Universality is sometimes expressed in a somehow defectively simple
way: some critical properties (the universal ones) of a system are
given by space dimensionality and the local properties (i.e. near the
identity element) of the coset space $\cal G/\cal H$, where $\cal G$
is the symmetry group of the high temperature phase and $\cal H$ is
the remaining symmetry group of the broken phase (low temperature). As
we shall discuss, the subtle point making the above statement not
straightforward to use, is that $\cal G$ needs not to be the symmetry
group of the microscopic Hamiltonian, but that of the coarse-grained
fixed-point action.

On the spirit of the above statement, some time ago~\cite{AZARIA} a
seemingly complete classification was obtained of the Universality
Classes of three dimensional systems where $\cal G$=O(3).  In this
picture, a phase transition of a vector model, with O(3) global
symmetry and with an O(2) low temperature phase symmetry, in three
dimensions must belong to the O(3)/O(2) scheme of symmetry breaking
(classical Heisenberg model). In addition, if $\cal H$=O(1)=Z$_2$ is
the remaining symmetry, the corresponding scheme should be O(4)/O(3)
which is locally isomorphic to O(3)/O(1).\footnote{This statement
assumes that the {\em global} properties of the coset $\cal G/\cal H$
are irrelevant, only the local properties matter.}
This classification has been challenged by the chiral
models\cite{KAWAMURA}. However, the situation is still hotly debated:
some authors believe that the chiral transitions are weakly
first-order\cite{NOCHIRAL}, while others claim\cite{ChiralN23} that
the Chiral Universality Class exists, implying the relevance of the
global properties of $\cal G/\cal H$.

In this letter, we shall consider the three dimensional
antiferromagnetic (AFM) $\RP2$
model~\cite{RP2D3LETTER,RP2D3LONG,ROMANO,SHROCK}, a model displaying a
second-order phase transition and escaping from the previously
expressed paradigm. It is worth recalling~\cite{PRBraiz,raiz} that one
of the phase transitions found in models for colossal
magnetorresistance oxides~\cite{CMR} belongs to the Universality class
of the AFM $\RP2$ model. The microscopic Hamiltonian of this model has
a global O(3) symmetry group, while the low temperature phase has, at
least, a remaining O(2) symmetry~\cite{PRBraiz}. We will show here
that the model belongs to the Universality Class of the three
dimensional O(5) Non Linear $\sigma$ model. Some ground for this
arises from a hand-waving argument, suggested to us by one of the
referees of Ref.~\cite{PRBraiz} (see below).

The Universality Class of the three dimensional O(5) Non Linear
$\sigma$ model has received less attention that $\OO{N}$ models with
$N=0,1,2,3$ and $4$. In spite of that, it has been recently argued
that $\OO5$ could be relevant for the High-temperature superconducting
cuprates~\cite{SO5}. Nevertheless, perturbative field
theoretic-methods have been used to estimate the critical
exponents~\cite{omega1,ANTONENKO,KLEINERT,BUTTI}. From the numerical
side, only a rather unconvincing Monte Carlo simulation~\cite{PRL-HU}
was available until very recently. Fortunately, there has been a
recent, much more careful study~\cite{PELISSETTOO5}. Yet, the scope of
Ref.~\cite{PELISSETTOO5} was to determine whether an interaction
explicitly degrading the $\OO5$ symmetry to an $\OO3\oplus\OO2$ group was
relevant in the Renormalization-Group sense. To that end, those
authors concentrated in producing extremely accurate data on small
lattices.

Our purpose is to study in greater detail the critical properties of
the three dimensional O(5) Non Linear $\sigma$ model, and of the AFM
$\RP2$ model. We improve over previous studies of both models,
obtaining more accurate estimates for critical exponents, universal
dimensionless quantities and non universal critical couplings. As
symmetries play such a prominent role, we will also explore the
possibilities of changing those of the low-temperature phase by adding
a second-neighbors coupling to the Hamiltonian of the AFM $\RP2$
model.

\section{The models}\label{MODELSECT}

We are considering a system of $N$-component normalized spins
$\{\vec{v_i}\}$ placed in a three dimensional simple cubic lattice of
size \textit{L} with periodic boundary conditions. The actions of our
lattice systems are

\begin{equation}
{\cal S}^{\OO{N}}=-\beta \sum_{<i,j>}\left(\vec{v_i} \cdot \vec{v_j}
\right) \ , \qquad {\cal S}^{\mathbf{R}\mathrm{P}^{N-1}}=-\beta
\sum_{<i,j>}\left(\vec{v_i} \cdot \vec{v_j} \right)^2\ ,
\label{accion}
\end{equation}
where the sums are extended to all pairs of nearest neighbors. Our
sign convention is fixed by the partition function:
\begin{equation}
Z=\int\,\prod_i d\vec v_i\ \e^{-{\cal S}}\ ,
\end{equation}
$d\vec v\/$ being the rotationally invariant measure over the
$N$-dimensional unit sphere.

To construct observables, in addition to the vector field $\vec{v_i}$,
we consider the (traceless) tensorial field
\begin{equation}
\tau_i^{\alpha\beta}=v_i^\alpha v_i^\beta-\frac1N\delta^{\alpha\beta}\ ,\qquad
\alpha,\beta=1,\ldots,N\ .
\end{equation}
The interesting quantities related with the order parameters can be
constructed in terms of the Fourier transforms of the fields
($f_i=\vec v_i,\tau_i$)
\begin{equation}
\hat{f}(\bbox{p})=\frac{1}{L^3}
\sum_i \e^{-\ii \bboxs{p}\cdot\bboxs{r}_i} f_i\ .
\end{equation} 

For $\RP{N-1}$ models, the local gauge invariance $\vec v_i\to-\vec
v_i$ implies that the relevant observables are constructed in terms of
the tensor field. However, for $\OO{N}$ we have found very interesting to
consider as well quantities related with the tensor field.

We construct the scalars (under global $\OO{N}$ transformations and
spatial translations)
\begin{equation}
S_\mathrm{V}(\bbox{p})=\hat{\vec v}(\bbox{p})\cdot\hat{\vec v}^*(\bbox{p}) ,\qquad 
S_\mathrm{T}(\bbox{p})=\mathrm{tr}\ \hat{\tau}(\bbox{p}) \hat{\tau}^*(\bbox{p})\ ,
\end{equation}
which, in addition to the action, are the only quantities measured
during the simulation. Their mean values yield the
propagators:
\begin{equation}
G_{T,V}(\bbox{p})=L^3\langle S_{T,V}(\bbox{p})\rangle\ .
\end{equation}

In the thermodynamic limit and at the critical point, the propagator
is expected to have poles at $\bbox{p}^\mathrm{f}_0=(0,0,0)$ and, for
the antiferromagnetic model, at
$\bbox{p}^\mathrm{s}_0=(\pi,\pi,\pi)$:\footnote{In the remaining part
of this section, if a subindex $V$ or $T$ does not explicitly appears,
it will imply that the equation is valid {\em both} for vector or
tensor quantities. The same convention will apply for superscript (f)
(ferromagnetic) and (s) (staggered). Staggered quantities are useful
only for antiferromagnetic models.}
\begin{equation}
G(\bbox{p}_0+\delta \bbox{p})\approx
\frac{Z\xi^{-\eta}}{(\delta \bbox{p})^2+\xi^{-2}}\,,
\end{equation}
where $\xi\Vert\delta \bbox{p}\Vert\ll 1$, and the exponents $\eta^{\f}$ and
$\eta^{\s}$ correspond to independent wave function renormalization at
each pole. Note that close to the critical point $\xi^\mathrm{f}$ and
$\xi^\mathrm{s}$ are expected to remain proportional to each other
(this will be explicitly checked numerically).

The (non-connected) susceptibilities are simply:
\begin{equation}
\chi = G(\bbox{p}_0)\ .
\end{equation}

In a finite lattice an extremely useful definition of the correlation
length can be obtained from the (discrete) derivative of
$G(\bbox{p})$.  Using $\delta \bbox{p}=(2\pi/L,0,0)$ one
obtains\cite{COOPER,AMIT}
\begin{equation}
\xi=\left(\frac{G(\bbox{p}_0)/G(\bbox{p}_0+\delta \bbox{p})-1}{4\sin^2(\pi/L)}\right)^{1/2}\ .
\end{equation} 
We also compute the cumulants
\begin{equation}
U_4=\frac{\langle S^2\rangle}{\langle S\rangle^2}\ .
\end{equation}
Finally, the energy per link is 
\begin{equation}
E=\langle {\cal S}/(-3 \beta L^3) \rangle\ .
\end{equation}

We have computed $\beta$-derivatives of observables through their
connected correlation with the action. Furthermore, we have
extrapolated mean-values from the simulation coupling, to a
neighboring value of $\beta$ using the standard reweighting
techniques, that cover all the relevant part of the critical
region~\cite{AMIT}.

The relationship between the $\OO5$ model and the $\RP2$ model arises
from the Landau-Wilson-Fisher Hamiltonian for the $\RP2$
system~\cite{PRBraiz}. Indeed, at the Mean Field
level~\cite{PRBraiz,RP2D3LONG}, the ferromagnetic quantities are
simple functions of the  staggered ones.  This suggests to
construct the Landau-Wilson-Fisher Hamiltonian from the staggered
magnetization, which is a traceless, real, symmetric $3\times 3$
matrix:
\begin{equation}
{\cal M}_\s^{\alpha,\beta}= \sum_i \ (-1)^{x_i+y_i+z_i}\tau_i^{\alpha,\beta}\,.
\end{equation}
Note that ${\cal M}_\s$ has 5 independent quantities. It is therefore
a simple matter to obtain a five-components real vector $\vec v$ such
that $\vec v^{\,2} =\mathrm{tr}\, {\cal M}_\s^2$. The less trivial part
regards the fourth-order interaction terms. In principle, the $\OO3$
symmetry of the microscopic Hamiltonian would allow for a $\mathrm{tr}
{\cal M}_\s^4$ term and a $[\mathrm{tr} {\cal M}_\s^2]^2$.
Surprisingly enough, both terms are proportional to 
$(\vec v^{\,2})^2$. Thus, assuming that sixth-order terms are irrelevant, the
Landau-Wilson-Fisher Hamiltonian is expected to have a $\OO5$ symmetry
group and both models belong to the same Universality Class. This does
not only implies that both models have the same critical exponents but
also that the $L\to\infty$ limit of $U_{4,\mathrm{V}}^{\OO5}$ and of
$U_{4}^{\s,\RP2}$ (evaluated at their respective critical couplings)
coincide.

\section{Numerical methods}\label{MONTECARLOSECT}

In the $\OO5$ model we have studied lattice sizes $L=6,8, 12, 16, 24,
32, 48, 64, 96$ and $L=128,$ at $\beta=1.1812$.  We have combined a
Wolff's single cluster update with Metropolis. Our {\em elementary\/}
Monte Carlo step (EMCS) consists of $(10 L +1)$ Wolff's cluster update
and then a full-lattice Metropolis sweep. We take measurements after
every EMCS.  Since the average size of clusters grows as
$L^{2-\eta}\approx L^2$, $80\%$ of simulation time we are tracing
clusters for all $L$, while Metropolis accounts for $10\%$ of time and
measurements for the remaining $10\%$.  The total simulation time time
has been the equivalent to 600 days of Pentium IV at 3.2 GHz. The
number of EMCS ranges from $10^8$ for $L=6$ to $1.4\times 10^6$ at
$L=128$. The integrated autocorrelation times for the susceptibility
and for the energy are smaller than 1 EMCS for all the simulated
lattice sizes.

Since we are interested in high accuracy estimates, we have used
double precision arithmetics. One also needs to worry about the pseudo
random-number generator. We have therefore implemented a
Schwinger-Dyson test. It turned out that the 32-bits Parisi-Rapuano
pseudo random-number generator~\cite{PARISI-RAPUANO} produces biased
results. Either the Parisi-Rapuano plus congruential
generator~\cite{SD} or the 64 bits Parisi-Rapuano generator cured this
bias. The 64 bits Parisi-Rapuano generator is faster and it has been
our final choice.

For the antiferromagnetic $\RP2$ model, no efficient cluster method is
available.  We have simulated in lattice sizes from $L=8, 12, 16, 24,
32, 48$ and $L=64$ at $\beta=-2.41$. We used a multi-hit Metropolis
sequential algorithm. Making a new spin proposal completely
independent from the previous spin value, we achieve an acceptance of
about $30\%$. We have used 2 hits what ensures a 50\% acceptance.  The
observables have been measured every two Metropolis full lattice-sweep
(our EMCS).

The number of EMCS ranges from $10^8$ for $L=8$ to $7\times 10^8$ for
$L=64$. In units of the integrated autocorrelation time $\tau$ (for
the order parameter) we have more than $10^6\tau$ for $L=64$. The data
up to $L=48$ were obtained in Pentium IV clusters (simulation time was
roughly equivalent to a 1000 days of a single processor).  For the
largest lattice, data were obtained in the {\em Mare Nostrum} computer
of the Barcelona Supercomputing Center (simulation time was roughly
equivalent to 3000 days of a single processor).

We perform a Finite-Size Scaling analysis, using the quotients
method~\cite{RP2D3LETTER,RP2D3LONG,AMIT}. In this approach, one
compares the mean value of an observable, $O$, in two systems of sizes
$L_1$ and $L_2$, at the value of $\beta$ where the correlation-length
in units of the lattice sizes coincides for both systems. If, for the
infinite volume system, $\langle O\rangle(\beta)\propto|\beta
-\beta_\mathrm{c}|^{-x_O}\,,$ the basic equation of the quotient
method is
\begin{equation}
Q_O^{L_1,L_2}\equiv\left.\frac{\langle O(\beta,L_2) \rangle}{\langle
    O(\beta,L_1)
    \rangle}\right|_{\frac{\xi(L_2,\beta)}{\xi(L_1,\beta)}
    =\frac{L_2}{L_1}}= \left( \frac{L_2}{L_1}
    \right)^{{x_O}/\nu}(1+A_OL_1^{-\omega}+\ldots)\,,
\label{QUOTIENTS-FORMULA}
\end{equation}
where the dots stand for higher order scaling corrections, $\nu$ is
the correlation-length critical exponent, $-\omega$ is the (universal)
first irrelevant critical exponent, while $A_O$ is a non universal
amplitude.  In a typical application, one fixes the ratio $s=L_2/L_1$
to 2, and consider pairs of lattices $L$ and $2L$. A linear
extrapolation in $L^{-\omega}$ is used to extract the infinite volume
limit. One just needs to make sure that the minimum lattice size
included in the extrapolation is large enough to safely neglect the
higher-order corrections. Of course, any quantity scaling like $\xi$
at the critical point, such as $LU_4$, may play the same role in
Eq.(\ref{QUOTIENTS-FORMULA}). However, usually $\xi$ yields smaller
scaling corrections than $U_4$.

The extrapolation method based on Eq.(\ref{QUOTIENTS-FORMULA}) is
feasible for the antiferromagnetic $\RP{2}$ model. Unfortunately, for
the $\OO5$ model the amplitude $A_O$ is surprisingly small. In fact,
resummation of the $\epsilon$-expansion yields
$\omega=0.79(2)$~\cite{omega1}, while blind use of
Eq.(\ref{QUOTIENTS-FORMULA}) on our numerical data would predict
$\omega\approx 2$. We have then considered an additional correction
term, $\tilde A_O L^{-\sigma}$. The exponent $\sigma$ is an effective
way of taking into account a variety of higher-order scaling
corrections of similar magnitude (an $L^{-2\omega}$ contribution,
subleading universal irrelevant critical corrections, analytic
corrections, effects of the non-linearity of the scaling fields,
etc.\cite{AMIT}).  Its utility will be in that it allow us to give
sensible error estimates for the infinite-volume extrapolations,
instead of bluntly taking $A_O=0$.

The most precise way of extracting the critical exponent, $\omega$, and
the critical point $\beta_\mathrm{c}$ is to consider the crossing point
of dimensionless quantities such as $\xi/L$ and $U_4$. Indeed, comparing
their values in lattices $L_1$ and $L_2$, one finds that they take
a common value at 
\begin{equation}
\beta^{L_2,L_1}_\mathrm{c}=\beta_\mathrm{c}+ B \frac{1 -
(L_2/L_1)^{-\omega}}{(L_2/L_1)^{1/\nu}-1} L_1^{-\omega -1/\nu} +
\ldots\,,\label{ajuste}
\end{equation}
The non universal amplitude $B$ depends on the considered dimensionless
quantity.  Again, one usually take pairs of lattices $L$ and $2 L$,
and extrapolates to infinite volume using Eq.(\ref{ajuste}), maybe
performing a joint fit for the crossing points of several dimensionless
quantities. Again, for the $\OO5$ model the amplitudes $B$ are
exceedingly small, and we need to add to Eq.(\ref{ajuste}) an
analogous higher-order term, where $\sigma$ plays the role of $\omega$,
and with amplitude $\tilde B$. 

\section{Results for the $\OO5$ model}

\begin{table}[h]
\begin{center}
\begin{tabular}{|r|c|c||c|c|}
\hline
$L$ & $\beta_{\mathrm{c},\,\xi_\V/L}^{L,2L}$ & $\xi_\V/L$ &  $\beta_{\mathrm{c},\,U_{4,\V}}^{L,2L}$& $U_{4,\V}$ \\
\hline
\hline
6
&1.179331(10)&0.275961(19)& 1.182619(19)& 1.069593(17)  \\
\hline	     						      
8	     						      
&1.180656(8) &0.278757(18)& 1.181896(12)& 1.069544(16)  \\
\hline	     						      
12	     						      
&1.181202(4) &0.280487(17)&1.181492(7) & 1.069673(14)  \\
\hline	     						      
16	     						      
& 1.181313(4)&0.28105(2)  &1.181410(6) & 1.069705(18)  \\
\hline 	     						      
24	     						      
&1.181353(3) &0.28137(2)  & 1.181371(5)& 1.069746(17)  \\
\hline	     						      
32	     						      
&1.181365(3) &0.28146(3)  & 1.181371(4)& 1.06976(2)    \\
\hline	     						      
48	     						      
&1.181366(3) &0.28155(5)  &1.181373(5) & 1.06972(3)    \\
\hline	     						      
64	     						      
& 1.181362(4)&0.28142(9)  & 1.181370(6)& 1.06982(6)    \\
\hline
\end{tabular}
\end{center}
\caption{Effective critical points, $\beta_\mathrm{c}^{L,2L}$, of the
$\OO5$ model, from the crossing points of the dimensionless quantities
$\xi_\V/L$ and $U_{4,\V}$ obtained in lattices of sizes $L$ and $2L$. We also
show the value of both quantities at their respective crossing point.}
\label{cortesO5}
\end{table}

The first step is the location of the critical point and the scaling
corrections exponent. In table \ref{cortesO5} we show the crossing points
$\beta_\mathrm{c}^{L,2L}$ for the dimensionless quantities $\xi_\V/L$ and $U_{4,\V}$.
To study the finite size corrections, we need to fit them to
\begin{equation}
\beta_\mathrm{c}^{L,2L}\approx\beta_\mathrm{c}+B L^{-\omega-1/\nu} +\tilde B
L^{-\sigma-1/\nu}\ .\label{fitbetac}
\end{equation}
Fixing $\tilde B=0$ yields $\omega$ larger than 2 which is
unacceptable given the field theory estimate
$\omega=0.79(2)$\cite{omega1}. Our interpretation is that $B$ is too
small to be observed even with our 6-digit accuracy. We have therefore
fixed $\omega=0.79(4)$ and we have taken as fit parameters
$\beta_\mathrm{c}$, $B$, $\tilde B$ and $\sigma$. We have doubled the field
theory error in $\omega$ for safety. To further constrain
$\beta_\mathrm{c}$ (and $\sigma$), we have performed a joint fit of
the crossing points for both  $\xi_\V/L$ and $U_{4,\V}$, with the same
$\beta_\mathrm{c}$. For this model we always fit for $L\geq 8\,$.  The
results are (in all fits reported in this work, the full covariance
matrix was used):
\begin{equation}
\beta_\mathrm{c}=1.1813654(19)\ ,\quad\sigma=2.21(17)\ ,\quad\chi^2/\mathrm{dof}=7.1/8\ . \label{betac}
\end{equation}
Notice that, for using equation (\ref{fitbetac}), an estimate of $\nu$ is
needed.  Fortunately a rough estimate $\nu\approx 0.78$ (see below) is
enough, given the uncertainty in $\omega$ and $\sigma$. As for the
amplitudes of the leading correction term, we find
\begin{equation}
B_{\xi_\V/L}=0.004(6) \ ,\quad B_{U_{4,\V}}=-0.001(3) \ , 
\end{equation}
while the amplitudes $\tilde B$ are of order one. We then see that the
$L^{-\sigma}$ term is crucial in order to obtain a sensible error
estimate in the $L\to\infty$ extrapolation.

At this point we may obtain two universal quantities, $U_{4,\V}^*$ and
$\xi_\V^*/L$, namely the $L\to\infty$ limit of $U_{4,\V}$ and $\xi_\V/L$
evaluated {\em exactly} at the critical coupling. Again, due to the
smallness of the leading scaling-corrections, we extrapolated to
$L\to\infty$ using the following functional forms:
\begin{equation}
U_{4,\V}(\beta_\mathrm{c}^{L,2L},L)\approx U_{4,\V}^* + C_{U_{4,\V}} L^{-\omega} +\tilde
C_{U_{4,\V}} L^{-\sigma}\,,\label{EXTRAU4}
\end{equation}
and
\begin{equation}
\frac{\xi_\V(\beta_\mathrm{c}^{L,2L},L)}{L}\approx \frac{\xi_\V^*}{L} + C_{\xi_\V/L} L^{-\omega} +\tilde
C_{\xi_\V/L} L^{-\sigma}\,.\label{EXTRAXIL}
\end{equation}
Our numerical estimates for $U_{4,\V}(\beta_\mathrm{c}^{L,2L},L)$ and
$\xi_\V(\beta_\mathrm{c}^{L,2L},L)/L$ are displayed in the third and
fourth columns of table \ref{cortesO5}. Although scaling-corrections
are tiny, they can be clearly observed.  Using
Eqs. (\ref{EXTRAU4},\ref{EXTRAXIL}) we obtain 
\begin{eqnarray}
\frac{\xi_\V^*}{L}=0.28145(13)\ (\frac{\chi^2}{\mathrm{dof}}=\frac{2.9}{3}),\ 
U_{4,\V}^*=1.06978(5)\ (\frac{\chi^2}{\mathrm{dof}}=\frac{2.8}{3})\,,
\end{eqnarray}
while the amplitudes of the leading scaling-corrections are
\begin{equation}
C_{\xi_\V/L}=0.002(3)\ ,\ C_{U_{4,\V}}=0.0002(6)\,.
\end{equation}

\begin{table}[h]
\begin{center}
\begin{tabular}{|l|c|c|c|}
\hline
L & $\nu$ & $\eta_\mathrm{V}$ & $\eta_T$\\
\hline
\hline
6
&0.7963(3) &0.04343(6)  & 1.3499(3)\\
\hline
8
&0.7894(4) & 0.03837(7) & 1.34254(12)\\
\hline
12
& 0.7849(4)& 0.03554(5) & 1.33780(10) \\
\hline
16
& 0.7833(4)& 0.03462(6)& 1.33589(11) \\
\hline 
24
& 0.7819(7)& 0.03430(10)& 1.33455(19)\\
\hline
32
& 0.7802(14)& 0.03396(16)& 1.3332(2) \\
\hline
48
& 0.7817(19) & 0.0339(2) & 1.3322(4) \\
\hline
64
& 0.781(4)  & 0.0341(4)& 1.3321(7)\\
\hline
\end{tabular}
\end{center}
\caption{$L$ dependent effective values of exponents
$\nu$,$\eta_\mathrm{V}$ and $\eta_T$ for the $\OO5$ model, calculated
from $Q_O^{L,2L}$.}
\label{nu_eta_table}
\end{table}

\begin{figure}
\begin{center}
\includegraphics[angle=270,width=0.9\linewidth]{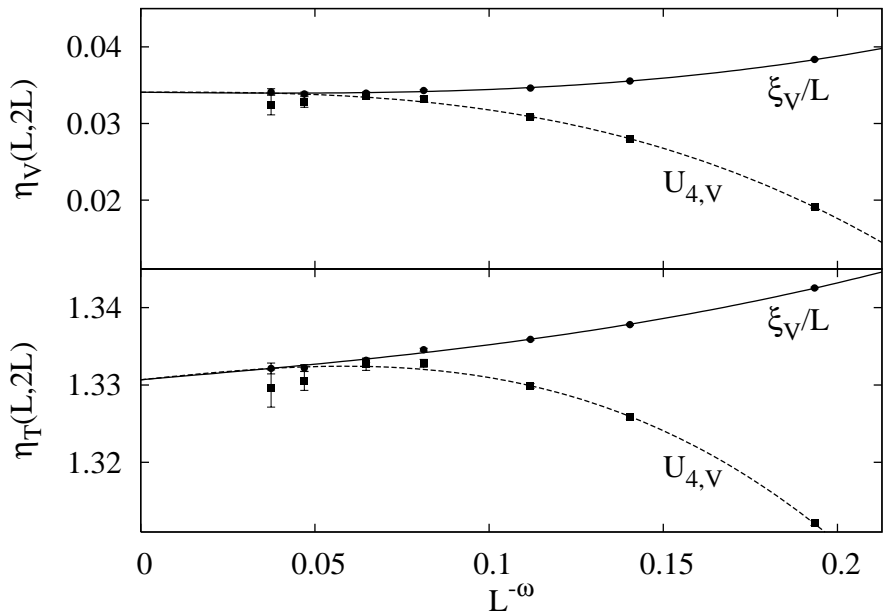}
\caption{Size-dependent estimators for the anomalous dimensions of the
$\OO5$ model as obtained from $(L,2L)$ pairs, versus $L^{-\omega}$. We
plot estimates from the crossing points of $U_{4,V}$ and of
$\xi_\V/L$.  We use the value $\omega=0.79$, from
Ref.\cite{omega1}.}\label{CORRECCIONES-O5}
\end{center}
\end{figure}

To obtain the critical exponents, we consider the operators
$\partial_\beta \xi_\V$ and $\chi_\mathrm{V,T}$ whose associated
exponents are $x_{\partial_\beta \xi_\V}=\nu+1$ and
$x_{\chi_\mathrm{V,T}}=\gamma_\mathrm{V,T}= \nu
(2-\eta_\mathrm{V,T})\,$. Taking the base 2 logarithm of the
quotients, see Eq.(\ref{QUOTIENTS-FORMULA}), we obtain the {\em
effective} size dependent exponents shown in table~\ref{nu_eta_table}.
In order to obtain their infinite volume value, we use
(\ref{QUOTIENTS-FORMULA}), including an explicit $L^{-\sigma}$ term in
the fit:
\begin{equation}
Q_O^{L,2L}= 2^{x_O/\nu} +A_O L^{-\omega}
    +\tilde A_O L^{-\sigma}\label{EXTRAEXPO}
\end{equation}
(note that we have absorbed a constant factor $2^{x_O/\nu}$ into the
amplitudes $A$ for scaling-corrections). We obtain 
\begin{equation}
\begin{array}{lllll}
\!\!\nu&=0.780(2), &\sigma=2.15(19), 
&A_{\partial_\beta \xi_\V}=-0.04(7),&\chi^2/\mathrm{dof}=8.4/8 \\
\!\!\eta_\mathrm{V}&=0.03405(3), &\sigma=2.27(19),
&A_{\chi_\mathrm{V}}=0.0012(14),&\chi^2/\mathrm{dof}=8.5/8\\
\!\!\eta_T&=1.3307(5),&\sigma=2.24(19),
&A_{\chi_\mathrm{T}}=-0.0053(12),&\chi^2/\mathrm{dof}=9.6/8
\end{array}\label{FinalO5}
\end{equation} 
Less than a $5\%$ of the total error is due to the error in $\omega=0.79(4)$ for both $\nu$ and $\eta_\V$.
For $\eta_T$ it is about a $30\%$.
There are two points to be made about the
extrapolation:
\begin{itemize} 
\item The $\OO5$ model is {\em not} an
improved action~\cite{PERFECT} (in the sense of exactly vanishing
leading scaling-corrections), since $A_{\chi_\mathrm{T}}$ is clearly
non-zero (this is also illustrated in Fig.~\ref{CORRECCIONES-O5}). Had we
not considered the tensorial operators, this would have been completely missed.
\item It is somehow disappointing to compare the accuracy of the
effective exponent $\nu$ in table \ref{nu_eta_table} with the error in
the extrapolation (\ref{FinalO5}). Indeed, could we safely set
$A_{\partial_\beta \xi_\V}=0$, the final result would have been $\nu =
0.7813(4)$.
\end{itemize}

\section{Results for the antiferromagnetic $\RP2$ model}

As we said before, qualitative arguments suggest that the
antiferromagnetic $\RP2$ model belongs to the $\OO5$ Universality
Class.  Our aim is to make the most astringent possible test of this
hypothesis, thus we perform here an update of a previous
study~\cite{RP2D3LONG} of the $\RP2$ critical quantities. We report
here largely improved estimates for critical coupling and
exponents. Furthermore, we give estimates for the dimensionless
quantity $U_{4}^{\mathrm{s},*}$, that can be directly compared with
the $U_{4,\V}^*$ obtained for the $\OO5$ model.

\begin{figure}
\begin{center}
\includegraphics[angle=270,width=0.9\linewidth]{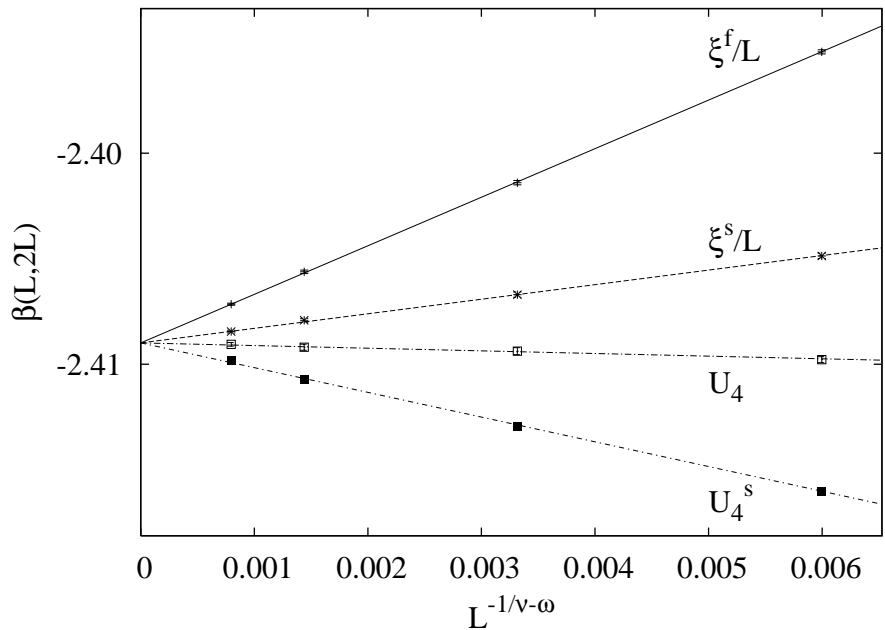}
\caption{Crossing points of the different dimensionless quantities,
for the  $(L,2L)$ pairs, as a function of $L^{-\omega -1/\nu}$. Lines are
fits to Eq.(\ref{ajuste}), constrained to yield common values of
$\beta_\mathrm{c}$ and $\omega$. The $\omega$ used in the $X$ axis
correspond to the optimal value. Note that the minimum pair plotted is
(12,24).}
\label{beta_c_rp2}
\end{center}
\end{figure}

In this case, the extrapolation to the infinite volume limit is more
standard (see for instance \cite{ISPERC}) than for the $\OO5$ model,
because the amplitude of the leading scaling-corrections are much
larger in most cases. To estimate $\omega$ and $\beta_\mathrm{c}$, we
consider pair of lattices $L$ and $2L$, performing a joint fit to
Eq.(\ref{ajuste}) of the crossing points of all four dimensionless
quantities, imposing a common value of $\beta_\mathrm{c}$ and $\omega$
(see Fig.~\ref{beta_c_rp2}).  To control for systematic errors due to
higher-order corrections, we follow the following procedure. We
perform the fit using data for $L\ge L_\mathrm{min}$, seeking a value
of $L_\mathrm{min}$ where a reasonable value of $\chi^2/\mathrm{dof}$ is
found. Furthermore, we require that the fit performed for $L >
L_\mathrm{min}$ yield compatible results. In that case, we report the
central value from the $L\ge L_\mathrm{min}$ fit, but taking the
enlarged errors from the $L > L_\mathrm{min}$ fit. We found that
$L_\mathrm{min}=12$ is enough for the extrapolation of $\beta_\mathrm{c}$.  We
obtain:
\begin{equation}
\beta_\mathrm{c}=-2.40899(13)\ ,\  \omega=0.78(4) \ ,\ \chi^2/\mathrm{dof}=8.5/10\,.\label{OMEGARP2}
\end{equation}

Once we have determined $\omega$, we proceed to extrapolate
$U_{4}^{\mathrm{s},*}$, and the critical exponents, using the analog
of Eqs.(\ref{EXTRAU4},\ref{EXTRAXIL},\ref{EXTRAEXPO}) without the
effective $L^{-\sigma}$ term. Although one could consider all four
types of crossing points, $\beta_\mathrm{c}^{L,2L}$, the resulting
quotients would be highly correlated, making join fits scarcely
useful. We concentrate on the crossing point of $\xi^\mathrm{s}/L$,
which seems the most natural quantity, as we are dealing with an
antiferromagnetic model and it can be obtained using only the two points
correlation function. We have checked that other choices for
$\beta_\mathrm{c}^{L,2L}$ yield compatible results, with slightly
larger errors. Our extrapolations are shown together with the
effective $L$ dependent estimates in
table~\ref{nu_eta_table_rp2}. Error estimates in the extrapolation
include the effect of the uncertainty in $\omega$.  For exponent $\nu$,
scaling corrections are completely buried in the statistical
errors. We extrapolated with a simple linear fit, using
$L_\mathrm{min}=8$. The situation is rather different for
$\eta^\s$. For that exponent, enlarging $L_\mathrm{min}$
systematically increases the asymptotic estimate. On the other hand,
a fit quadratic in $L^{-\omega}$ yields a linear term compatible with
zero. The linear extrapolation with $L_\mathrm{min}=16$ is identical
to the quadratic extrapolation from $L_\mathrm{min}=8$. This is the
result indicated in table~\ref{nu_eta_table_rp2}. As for $\eta^\f$, we
have rather strong leading scaling corrections. Indeed, a fit linear
in $L^{-\omega}$ yields basically identical results for
$L_\mathrm{min}=8$ and $L_\mathrm{min}=12$ (this is the result
reported in table~\ref{nu_eta_table_rp2}). Furthermore, a fit
quadratic in $L^{-\omega}$ including all points, yielded
$\eta^\f=1.331(5)$. The extrapolation for $U_4^{\mathrm{s},*}$ is
equally simple.

\begin{table}[h]
\begin{center}
\begin{tabular}{|c|c|c|c|c|c|}
\hline
$L$ & $\beta_{\mathrm{c},\,\xi^\mathrm{s}/L}^{L,2L}$ & $U_4^\mathrm{s}$ &
$\nu$ &$\eta^\mathrm{s}$& $\eta^\mathrm{f}$\\
\hline
\hline
8
&-2.39892(18) &1.06810(8) &0.7848(12)&0.0390(3) &1.4155(6)\\
\hline
12
&-2.40487(13)&1.06805(11) &0.7871(16)&0.0364(4) &1.3902(7)\\
\hline
16
&-2.40670(12) &1.06849(14) &0.782(2)&0.0357(6) &1.3776(11)\\
\hline
24
&-2.40792(9) &1.06883(19) &0.779(3)&0.0351(7) &1.3655(13) \\
\hline
32
&-2.40846(7) &1.06868(18) & 0.783(3)&0.0337(7) &1.3572(14)   \\
\hline
\hline
$\infty$
& -2.40899(13) &1.0691(5) & 0.780(4) & 0.032(2) &1.328(4)\\
\hline
\end{tabular}
\end{center}
\caption{Size dependent estimators for the critical coupling and
several Universal quantities, as obtained from $(L,2L)$ pairs, in the
$\RP2$ model. The last row correspond to the infinite volume
extrapolations.}\label{nu_eta_table_rp2}
\end{table}

The extrapolation for other scale-invariant quantities, without an
obvious correspondent in the $\OO5$ model, are:
\begin{equation}
\frac{\xi^{\s,*}}{L}=0.5379(17)\,,\ 
\frac{\xi^{\f,*}}{L}=0.2236(15)\,,\ 
U_4^{\f,*}=1.3114(6)\,.
\end{equation}

\section{Next nearest neighbors coupling}\label{NNNCOUPLING}

A rather subtle question regards the symmetry of the low-temperature
$\RP2$ antiferromagnetic phase~\cite{RP2D3LONG,PRBraiz}. A way of
investigating this problem is to study the enlarged action
\begin{equation}
{\cal S}=-\beta_1 \sum_{<i,j>}\left(\vec{v_i} \cdot \vec{v_j} \right)^2 -
\beta_2 \sum_{\ll i,j\gg}\left(\vec{v_i} \cdot \vec{v_j} \right)^2 \ .
\label{accion_rp2_2betas} 
\end{equation}
where an additional second-neighbors coupling is considered.

\begin{figure}
\begin{center}
\includegraphics[angle=270,width=0.9\linewidth]{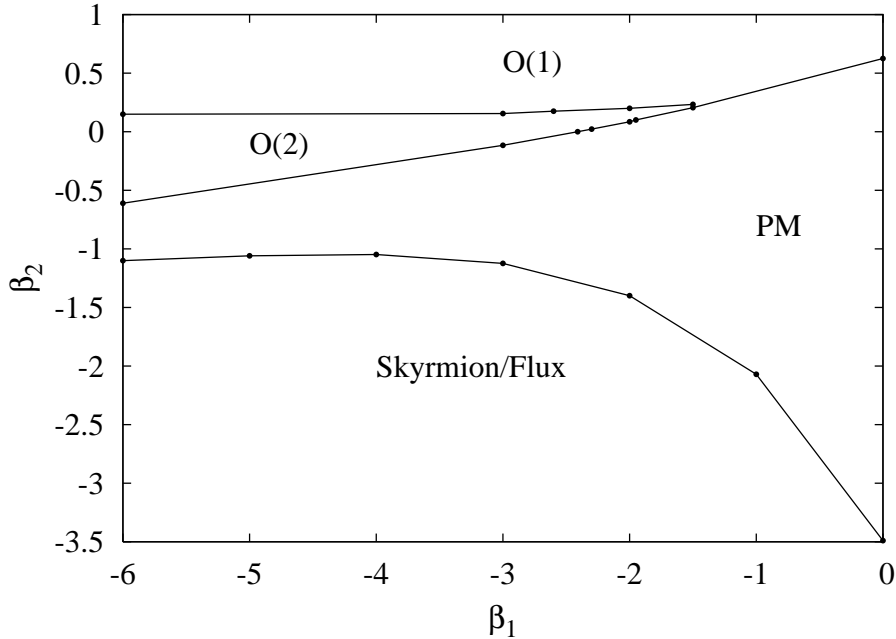}
\caption{Phase diagram of the extended antiferromagnetic $\RP2$ model,
Eq.(\ref{accion_rp2_2betas}).}
\label{diagrama_rp2}
\end{center}
\end{figure}

The phase diagram for $\beta_1 < 0$ (Fig. \ref{diagrama_rp2}) contains
the following regions (spins are classified as {\em even} or {\em
odd}, according to the parity of $x_i+y_i+z_i$):
\begin{itemize}
\item PM: the usual (paramagnetic) disordered state, where the $\OO3$
symmetry of the action (\ref{accion_rp2_2betas}) is preserved.
\item $\OO2$:  (say) even spins fluctuate almost parallel to (say)
the $Z$ axis, with random sense (local $Z_2$ symmetry), while odd spins
fluctuate in the perpendicular plane (global $\OO2$ symmetry)
\item $\OO1$: two sublattices with ferromagnetic ordering in
perpendicular directions, with random sense (the local $Z_2$ symmetry, $\vec{v_i}\rightarrow -\vec{v_i}$
is always preserved).
\item Skyrmion/Flux: the spins are parallel to the diagonals of the
unit cube, so that they point out from/to the center (i.e. the
propagator show three peaks at $\bbox{p}_0=(\pi,\pi,0)$ and
permutations). It is interesting to note the vectorial version of this
phase appear in models for colossal magnetorresistance
oxides~\cite{CMR-SKYRMIONS}.
\end{itemize}

The most relevant results can be summarized as follows:
\begin{itemize}
\item We have obtained critical exponents for several points along the
PM-$\OO2$ critical lines, with significantly less accuracy than for
the $\beta_2=0$ model. No variation was observed within errors.
\item The $\OO2$-$\OO1$ critical line is {\em repelled} from the
$\beta_2=0$ axis by the second-neighbors coupling. We interprete this
as a competition with the order-from-disorder
mechanism~\cite{RP2D3LONG,PRBraiz} behind the PM-$\OO2$ transition.
\item A naive analysis suggests that the $\OO2$-$\OO1$ transition line
should belong to the $XY$ universality class (consider first the limit
$\beta_1=-\infty$, then the identity $\cos^2\theta= (1+ \cos
2\theta)/2$ for the less ordered face-centered cubic
sublattice). However, we have found that this transition line is first
order, as revealed by the double-peaked histogram of the second
neighbors energy. We are able to estimate a non-zero latent-heat up
to $\beta_1=-6$. At $\beta_1=-4$ the double-peak structure is still
easy to observe on small lattices. We presume that the whole line
is first-order, although it could become very weak. 
\item The skyrmion-PM transition lines turned out to be first-order at all the checked points.
\item Note that at $\beta_1=0$, we have two decoupled ferromagnetic
$\RP2$ models on the face-centered cubic lattice (a model showing
first-order transition, well known in the liquids-crystal
context~\cite{LEWOHL_LASHER}). We should remark that a precise
location of the triple point $\OO2$-$\OO1$-PM is very difficult to achieve.
\end{itemize}

\section{Conclusions}

We have obtained high accuracy estimates of critical exponents and
other universal quantities for the three-dimensional $\OO5$ and the
antiferromagnetic $\RP2$ models, by means of Monte Carlo simulation,
Finite-Size Scaling analysis and careful infinite-volume
extrapolation.

In the case of the $\OO5$ model the coupling to the leading irrelevant
operator is rather weak, but non-vanishing, and one needs to consider
higher-order scaling corrections to obtain sensible error
estimates. In spite of that, our estimates for the critical coupling
and the anomalous dimensions for the vector and tensor representations
improve significantly over previous work (see table \ref{summary}).

For the $\RP2$ model, the leading scaling corrections are sizeable. It
is amusing that we are able to obtain numerically (for the first time,
we believe) an estimate for the (universal) scaling-correction
exponent $\omega=0.78(4)$, of accuracy comparable to the perturbative
field-theoretical estimate\cite{omega1} $\omega=0.79(2)$.  As table
\ref{summary} shows, within the achieved accuracy, both models seem to
belong to the same Universality Class (for comparison, we also show
results from the $\OO4$ Universality Class). To conclude this, one
needs to accept that in the $\RP2$ model, the wavefunction
renormalization for the propagator pole at $\bbox{p}_0=(\pi,\pi,\pi)$
is as for the the $\OO5$ fundamental field, while at
$\bbox{p}_0=(0,0,0)$ is as for the $\OO5$ tensor field.

We have also obtained the phase-diagram of the $\RP2$ model extended
with a second nearest-neighbors interaction. We have found a rich
phase diagram.

\begin{table}[h]
\scriptsize
\begin{center}
\begin{tabular}{|l|l|l|l|l|l|}
\hline
\multicolumn{1}{|c|}
{Model}&
\multicolumn{1}{c|}
{$\beta_{\mathrm{c}}$}&
\multicolumn{1}{c|}
{$\nu$}&
\multicolumn{1}{c|}
{$\eta$}&
\multicolumn{1}{c|}
{$\eta'$}&
\multicolumn{1}{c|}
{$U_4$}\\
\hline
\hline
$\RP2$ (this work)         
&$-2.40899(13)$&$0.780(4)$&$0.032(2)$&$1.328(4)$&$1.0691(5)$\\
\hline
$\OO5$ (this work)
&$+1.1813654(19)$&$0.780(2)$&$0.03405(3)$&$1.3307(5)$&$1.06978(5)$\\
\hline
$\OO5$ (Ref.~\cite{PELISSETTOO5})
&$+1.18138(3)$&$0.779(3)$&$0.034(1)$&---&1.069(1)\\
\hline
$\OO5$ (FT~\cite{omega1})
&---&$0.762(7)$&$0.034(4)$&---&---\\
\hline
$\OO4$ 
&$+0.935858(8)$\cite{ON}&$0.749(2)$\cite{HASENBUSCH}&$0.0365(10)$\cite{HASENBUSCH}&1.375(5)\cite{ON}&---\\
\hline
\end{tabular}
\end{center}
\caption{Summary of infinite-volume estimates for the 3D
antiferromagnetic $\RP2$, $\OO5$ and $\OO4$ models. We call $\eta'$ to
$\eta^\f$ for $\RP2$ and to the $\eta_\mathrm{T}$ for $\OO{N}$ models.
FT stands for Field-Theory.}\label{summary}
\end{table}

\section*{Acknowledgments}

We are indebted with Juan Jes\'us Ruiz-Lorenzo for discussions.  We are
grateful to the Barcelona Supercomputing Center, for allowing us to
use {\em Mare Nostrum} in its installation phase.  Isabel Campos was
very helpful in the handling of {\em Mare Nostrum}. About a third of
the total simulation time was obtained from the cluster of the
Instituto de Biocomputaci\'on y F\'{\i}sica de Sistemas Complejos, in
Zaragoza, and other computing resources of the Departamento de
F\'{\i}sica Te\'orica of Universidad de Zaragoza. We acknowledge
partial financial support from Ministerio de Educaci\'on y Ciencia
(Spain) though research contracts BFM2003-08532 and FIS2004-05073.

\end{document}